\documentclass{svjour2}
\usepackage{graphicx}  
\usepackage{latexsym}
\usepackage{amssymb}

\def\beq{\begin{equation}}
\def\eeq{\end{equation}}

\def\rmd{{\rm d}}

\journalname{General Relativity and Gravitation}

\begin{document}

\title{Separable geodesic action slicing in stationary spacetimes}

\author{Donato Bini\and
Andrea Geralico\and 
Robert T. Jantzen
}

\institute{
Donato Bini 
\at
Istituto per le Applicazioni del Calcolo ``M. Picone,'' CNR, I--00161 Rome, Italy\\
ICRA, University of Rome ``La Sapienza,'' I--00185 Rome, Italy\\
INFN - Sezione di Firenze, Polo Scientifico, Via Sansone 1, I--50019, Sesto Fiorentino (FI), Italy\\
INAF - Astronomical Observatory of Turin, Via Osservatorio 20, I--10025 Pino Torinese (TO), Italy\\
\email{binid@icra.it} 
\and
Andrea Geralico 
\at
Physics Department and ICRA, University of Rome ``La Sapienza,'' I--00185 Rome, Italy\\
\email{geralico@icra.it}   
\and
Robert T. Jantzen
\at
Department of Mathematics and Statistics, Villanova University, Villanova, PA 19085, USA\\
ICRA, University of Rome ``La Sapienza,'' I--00185 Rome, Italy\\
\email{robert.jantzen@villanova.edu} 
}

\date{Received: date / Accepted: date / Version: date }

\maketitle

\begin{abstract}
A simple observation about the action for geodesics in a stationary spacetime with separable geodesic equations leads to a natural class of  slicings of that spacetime whose orthogonal geodesic trajectories represent freely falling observers. The time coordinate function can then be taken to be the observer proper time, leading to a unit lapse function. This explains some of the properties of the original Painlev\'e-Gullstrand coordinates on the Schwarzschild spacetime and their generalization to the Kerr-Newman family of spacetimes, reproducible also locally for the G\"odel spacetime. 
For the static spherically symmetric case the slicing can be chosen to be intrinsically flat with spherically symmetric geodesic observers, leaving all the gravitational field information in the shift vector field.

\keywords{spacetime slicing, Painlev\'e-Gullstrand coordinates}
\PACS{04.20.Cv}
\end{abstract}

\section{Introduction}

Natural spacelike slicings of spacetimes characterized by special geometric properties are very helpful in elucidating the structure of those spacetimes, as advocated by Smarr and York \cite{smarr-york}, for example, who show how various choices of lapse function and shift vector field can simplify the spatial metric. The family of orthogonal trajectories to such slicings represent the world lines of a family of test observers (`fiducial observers') which then experience the spacetime geometry in $3+1$ form, with the future-pointing unit normal vector field as their 4-velocity field. The most familiar and useful slicings in a stationary spacetime are often associated with nongeodesic slicings which are accelerated in order to resist the pull of gravity and link to our mental image from Newtonian physics of points fixed in space. The zero-angular-momentum observers (ZAMOs) in the stationary axisymmetric spacetimes like in the Kerr family of black holes are the standard tool for decomposing spacetime quantities in those spacetimes \cite{MTW}, resisting the attraction towards the hole while being dragged along by its rotation relative to the Boyer-Lindquist coordinate grid. The shift vector field describes the motion of these fiducial observers with respect to the time lines anchored in the spatial coordinate grid, while the lapse function acts like a potential for the acceleration field characterizing those observers, which move orthogonally to the time coordinate hypersurfaces. However, these coordinates  have a singularity at the outer event horizon where the time coordinate slices go null and then timelike as one continues inside, where the ZAMOs no longer exist.

Painlev\'e-Gullstrand coordinates \cite{doran,cook,hamilton}, 
which exist in the Kerr and Kerr-Newman spacetimes generalizing those first found for the Schwarz\-schild spacetime \cite{Painleve21,Gullstrand22},
are instead associated with a unit lapse gauge slicing whose corresponding orthogonal fiducial observers are both stationary and geodesic, and  represent a stationary field of freely falling observers  whose adapted coordinates have desirable properties. First and foremost these new coordinates remain valid inside the outer horizon in these spacetimes,
leading to the terminology `horizon-penetrating coordinates,'
while for example, in the Schwarzschild case the intrinsic geometry of the time slices is indeed flat. Retaining the original spatial coordinate functions (and therefore the same time coordinate lines adapted to the Killing vector associated with the stationary symmetry), passing to the new time function simply changes the fiducial observers used to interpret spacetime quantities, with an additional contribution to the shift to represent their motion with respect to the time lines. In the Kerr-Newman case, an additional change of azimuthal coordinate is then possible which drags the coordinate along by the geodesic motion in that direction. A similar situation occurs locally for the G\"odel spacetime, leading to a new analogous representation of that metric.

The definition of such new time and space coordinates can be explained by examining the action for the timelike geodesics of these spacetimes, which are special in the sense that the action itself and the
affinely parametrized geodesic equations are separable. In fact the geodesics are completely integrable, and in the Kerr case are determined by four first integrals related to that separability property which is due to the existence of Killing vectors associated with the stationary axisymmetric symmetry and the existence of a Killing tensor \cite{carter,MTW}. Of these four constants of the motion for a given geodesic, one can be absorbed into the choice of parametrization for the affinely parametrized geodesics, leaving three constants to determine the direction of their 4-velocity in spacetime. By fixing these constants for the entire 4-parameter spacetime, one determines a geodesic congruence which is vorticity free, and admits a family of orthogonal spacelike hypersurfaces  serving as the time slices of a useful coordinate system adapted to this family in a unit lapse time gauge. These choices of the constants of the motion are further limited in order to avoid limiting the range of validity of the new coordinates by energy or angular momentum barriers. A similar situation holds for the G\"odel spacetime due to its stationary cylindrical symmetry when expressed in cylindrical-like coordinates.

The new time slices are closely related to the hypersurfaces of constant action for the geodesic problem because of the separability property, and their parametrization measures the proper time along the geodesics. This fact unifies the derivation of the various examples of unit lapse time gauge coordinates generalizing the original Painlev\'e-Gullstrand coordinates found for the Schwarzschild spacetime. For that spacetime in addition to unit lapse, the new coordinates have intrinsically flat time coordinate slicings, a property which is another route to generalize the Painlev\'e-Gullstrand coordinates with accelerated observers, as done for  the de Sitter spacetime \cite{parikh}.  In that case as well as the anti de Sitter case, in a region which admits a static spherically symmetric slicing, one can also introduce unit lapse gauge slicings with geodesic fiducial observers following from this separability discussion.
In the static spherically symmetric case, the geodesic slicing corresponding to purely radial motion of the observer family can always be chosen to be intrinsically flat. One motivation for considering these kinds of coordinate systems comes from numerical relativity where horizon-penetrating coordinates like these can be useful  \cite{cook,gourgo}.

\section{Separable geodesic action slicings}

Spacetimes with sufficient Killing symmetries allow coordinates to be introduced which allow the separation of variables for the geodesic equations \cite{carter,woodhouse,dietz,collinson1,collinson2,demianski,koutras,houri2008,houri2011}. Carter \cite{carter} was the first to appreciate this fact and determine a large class of exact solution stationary axisymmetric spacetimes with this property.   

Let the coordinates $x^\alpha$ ($\alpha=0\ldots 3$, with $x^0=t$) be such that the geodesic equations are separable in the metric 
$\rmd s^2=g_{\alpha\beta}\rmd x^\alpha \rmd x^\beta$
of signature $-$$+$$+$$+$.
Using the Hamilton-Jacobi formalism we can write the tangent vector $U^\alpha=\rmd x^\alpha(\lambda)/\rmd\lambda$ to the affinely parametrized timelike geodesics
as the gradient of the fundamental action function $S=S(x^\alpha,\lambda)$, $U_\alpha=\partial_\alpha S$, satisfying
the Hamilton-Jacobi equation
\beq
-\frac{\partial S}{\partial \lambda}=H(x^\alpha,\partial_\alpha S)\,,
\eeq
with $\lambda$ an affine parameter for the integral curves of $U$ and the Hamiltonian
\beq
H=\frac12 g^{\alpha \beta}\partial_\alpha S \partial_\beta S=-\frac12\mu^2={\it const}\,,
\eeq
the latter identity following from the normalization condition $U^\alpha  U_\alpha =-\mu^2$ for timelike geodesics. The choice $\mu=1$ makes $\lambda$ equal to the proper time along the geodesics and $U^\alpha$ the unit 4-velocity \footnote{The present analysis can be easily generalized to the case of charged background spacetimes and nongeodesic orbits  of charged particles (with charge $e$) still allowing a separable action and Hamilton-Jacobi equation
\beq
g^{\alpha\beta}(\partial_\alpha S -eA_\alpha)(\partial_\beta S -eA_\beta)=-\mu^2
\eeq
as occurs for a Kerr-Newman black hole. We will not discuss such a generalization here.
}.

Assume that $S$ can be separated in its dependence on the variables $x^\alpha$ and $\lambda$, namely
\beq
S=\frac12\mu^2\lambda+S_t(t)+S_1(x^1)+S_2(x^2)+S_3(x^3)\,. 
\eeq 
Thus we have for the 1-form $U^\flat\equiv U_\alpha \rmd x^\alpha =\partial_\alpha S \rmd x^\alpha=\rmd (S-\frac12\mu^2 \lambda)$, where here $\rmd$ stands for the spacetime differential only, or explicitly
\beq
U^\flat  = \partial_t S_t(t) \rmd t+ \partial_1 S_1(x^1) \rmd x^1+ \partial_2 S_2(x^2) \rmd x^2+ \partial_3 S_3(x^3) \rmd x^3\,.
\eeq
Moreover, since in this case $U$ is a gradient it is also necessarily vorticity-free: 
$\rmd U^\flat=0$,
and there exists a distribution of constant action hypersurfaces $T\equiv -S+\frac12\mu^2 \lambda =const$ with
\beq
\label{eq:T1}
-\rmd T=U_\alpha \rmd x^\alpha\,,
\eeq
such that $U^\alpha$ is the associated unit normal vector field. When one sets $\mu=1$, then the time function $T$ measures the proper time along the geodesics and the corresponding lapse function has the fixed value $N=1$. For a stationary spacetime in which $t$ is taken to be a Killing time coordinate, then $U_t = -E$ is a constant interpreted as a conserved energy, with $S_t(t)=-Et$, and the metric is independent of $t$. One then has
\beq
\label{eq:T}
-\rmd T=-E\rmd t + U_a \rmd x^a\,.
\eeq

In the Schwarzschild case following this procedure starting from the usual Boyer-Lindquist coordinates, one not only simplifies the lapse function but also the spatial metric, which becomes flat for certain choices of the parameters, leaving the shift vector field to carry the information about the gravitational field. To study the intrinsic geometry one needs to evaluate the induced metric on this new slicing, whose $3$-dimensional line element will be denoted by ${}^{(3)}\rmd s^2=\gamma_{ab}\rmd x^a \rmd x^b$. 

When $\partial_t$ is a timelike Killing vector the spacetime metric coefficients $g_{\alpha\beta}$ do not depend on $t$.
Moreover, from Eq.~(\ref{eq:T}), on the $T=\hbox{\it const}$ slicings  we have
\beq
\label{dt_su_T}
\rmd t= \frac{U_a}{E} \rmd x^a\,,\quad (a=1,2,3)\,.
\eeq
This represents a tilting of the original slicing tangent spaces to adapt them to the new stationary geodesic observer family.
Substituting the above expression (\ref{dt_su_T}) into the metric we then have
\begin{eqnarray}
{}^{(3)}\rmd s^2&=& 
(g_{tt}\rmd t^2 +2g_{ta}\rmd t \rmd x^a +g_{ab}\rmd x^a \rmd x^b)|_{dt=dx^a \,U_a/E}
\nonumber \\
&=& g_{tt}\frac{U_aU_b}{E^2} \rmd x^a\rmd x^b  +2g_{ta}\frac{U_b}{E} \rmd x^a\rmd x^b  +g_{ab}\rmd x^a \rmd x^b
\equiv\gamma_{ab}\rmd x^a\rmd x^b
\end{eqnarray}
where the spatial metric components in the original spatial coordinates
\beq\label{eq:gamma}
\gamma_{ab}=g_{tt}\frac{U_aU_b}{E^2}   +\frac{2}{E}g_{t(a} U_{b)} +g_{ab}\,,
\eeq
are also independent of $t$ since the time coordinate lines are still Killing trajectories.

In the unit lapse time gauge $N=1$ of these new coordinates, the components of the shift vector field $N^a = N^2 g^{Ta} = g^{Ta}$
are given simply by the mixed components of the contravariant metric tensor \cite{mfg}. These describe the motion of the Killing time lines relative to the new fiducial observers. Although the relative velocity of these time lines with respect to these geodesic observers exceeds the speed of light within the horizon, the new coordinates remain well defined. The spacetime metric is then
\beq
 \rmd s^2=- \rmd T^2 +\gamma_{ab} (\rmd x^a+N^a\rmd T)  (\rmd x^b+N^b \rmd T)\,.
\eeq

\section{Static spherically symmetric spacetimes}
\label{stat}

Before considering the more complicated case of stationary axisymmetric spacetimes like the Kerr spacetime, consider first the 
static spherically symmetric spacetimes. The metric written in standard spherical-like coordinates is
\begin{equation}
\label{metricgen}
\rmd s^2=- e^{\nu}\rmd t^2 + e^{\lambda}\rmd r^2+r^2(\rmd \theta^2 +\sin ^2\theta \rmd \phi^2)\,,
\end{equation}
where the functions $\nu$ and $\lambda$ depend only on the radial coordinate. Then $L=U_\phi$ is an additional Killing constant associated with the conserved angular momentum so
\beq
U_\alpha \rmd x^\alpha = -E\rmd t + (\partial_r S_r) \rmd r +(\partial_\theta S_\theta) \rmd \theta+L\rmd \phi \,,
\eeq
and the corresponding Hamilton-Jacobi equation 
\beq
- e^{-\nu} E^2 + e^{-\lambda} (\partial_r S_r)^2 +\frac{1}{r^2}\left[(\partial_\theta S_\theta)^2 +\frac{L^2}{\sin^2\theta} \right]
 =-\mu^2
\eeq
can be easily separated in its dependence on the coordinates 
leading to
\beq
 \frac{\rmd S_r}{\rmd r} =\epsilon_r e^{\lambda/2}\sqrt{E^2e^{-\nu}-\frac{K +\mu^2 r^2}{r^2}}\,,\qquad
 \frac{\rmd S_\theta}{\rmd \theta} =\epsilon_\theta  \sqrt{K-\frac{L^2}{\sin^2\theta}}\,,
\eeq
where $K$ is a separation constant arising from the angular contribution to the Hamilton-Jacobi equation, and
$|\epsilon_r|=1=|\epsilon_\theta|$.

As stated above, we can set $\mu=1$ to characterize a new foliation by a new temporal coordinate $T$ measuring proper time along the orthogonal geodesics.
We are left to specify $E$, $L$, and $K$ to obtain a specific family of timelike geodesics covering the spacetime.
The simplest choice would be a spherically symmetric 4-velocity field involving only radial motion of the geodesics relative to the original coordinates. We can achieve this in two steps. First we can require that this family of geodesics be tangent to the equatorial plane $\theta=\pi/2$, which requires $K=L^2$ to make $U_\theta=0$, resulting in 
\beq
U_r^2=e^{\lambda-\nu}\left[E^2-e^{\nu}\left(1+\frac{L^2}{r^2}\right)\right]\,.
\eeq 
We then impose the radial condition $L=0$, so that
\beq\label{eq:Ur}
U_r = \epsilon_r e^{(\lambda-\nu)/2}\sqrt{E^2-e^{\nu}} \,,
\eeq
leaving finally the choice of the energy constant $E$. For spatially asymptotically flat spacetimes where $e^{\nu}<1$ approaches 1 as $r\to\infty$,
to have a choice which works even at spatial infinity, we must have $E\ge 1$, in which case the value may be interpreted as the energy of the radially moving geodesics at spatial infinity. Of course one could choose $E<1$ but this would limit the slicing to the interior of a cylinder in spacetime inside the radial turning point of the geodesic motion.

The new time differential is then
\beq
\label{Tpglike}
\rmd T=E\rmd t -U_r\, \rmd r  \,.
\eeq
A new global coordinate system for static spacetimes is given by $(X^\alpha)=(T,R,\theta,\phi)$ with $R=r$ and $\theta$ and $\phi$ unchanged and $T=Et+f(r)$ given by integrating the differential equation
$f'(r) =- e^\lambda U_r$. This leads to
\beq
   \partial_T = {E}^{-1}  \partial_t\,,\quad
   \partial_R =  \partial_r +U_r  \partial_t\,,
\eeq 
and
the transformed metric is 
\beq
\label{newmet}
\rmd s^2=-\rmd T^2+\gamma_{ab}(\rmd X^a+N^a\rmd T)(\rmd X^b+N^b\rmd T)\,,
\eeq
with unit lapse function and the shift vector field aligned with the new radial direction, i.e.,
\beq
N^a=-\delta^a_R \epsilon_r e^{-\lambda} U_r
  =  - \delta^a_R \epsilon_r e^{-(\lambda+\nu)/2}\sqrt{E^2-e^{\nu}}
\,.
\eeq
The 3-metric induced on the $T=\hbox{\it const}$ hypersurfaces is then given by
\beq
\label{3_metric}
{}^{(3)}\rmd s^2=\frac{e^{\lambda+\nu}}{E^2}\rmd r ^2+r^2(\rmd \theta^2+\sin^2\theta\rmd \phi^2)\,.
\eeq

In the case of vacuum as well as in the presence of a nonzero cosmological constant one has $\lambda+\nu=0$, so that the induced metric is  then
\beq
\label{indmetE}
  {}^{(3)}\rmd  s^2 =\frac{\rmd r^2}{E^2}+ r^2 (\rmd \theta^2 +\sin^2 \theta \rmd \phi^2)\,,
\eeq
whose only nonvanishing component of the spatial Riemann curvature tensor and the spatial curvature scalar are
\beq
  {}^{(3)} R^{\theta\phi}{}_{\theta\phi} = \frac{1-E^2}{r^2} =\frac12 {}^{(3)} R 
  \,,
\eeq
with positive or negative curvature respectively for $0<E<1 $ (bound geodesics) or $E>1$ (unbound geodesics).
The choice $E=1$ leads to a flat $3$-geometry. The additional sign choice $\epsilon_r=-1$ corresponds to the radially infalling geodesics which start at rest at spatial infinity. 
This is the case for the Schwarzschild spacetime where the Painlev\'e-Gullstrand coordinates were originally found.
For more details on flat foliations of spherically symmetric spacetimes see Refs. \cite{guven,beig,herrero}.

If one does not choose $L=0$, then an angular momentum barrier where $U_r^2$ changes sign prevents the new slicing from reaching the horizon in a way complementary to the way the choice $E<1$ prevents the slicing from reaching spatial infinity. Thus modulo the choice of sign for incoming or outgoing radial geodesics, the $L=0, E\ge1$ slicings are the only ones which cover the original region exterior to the horizon in the Schwarzschild case, as well as extending through that horizon
up to the singularity at $r=0$.

For other examples, one can specialize these general results to the case of the de Sitter and anti de Sitter spacetimes, which in static coordinate systems have the same metric function expressions
\beq
e^\nu=1-\frac{\Lambda}{3}r^2=e^{-\lambda}\,.
\eeq
but for the two different signs of $\Lambda$, respectively positive and negative. In the de Sitter case 
$\Lambda>0$,
the original coordinates are limited by a coordinate singularity at the radius at which this expression goes to zero.
Expressing the cosmological constant in terms of Hubble parameter $H$ in that case we have $\Lambda=3H^2$ and the metric (\ref{metricgen}) in the original coordinates only covers the region $0<r<1/H$.
Timelike geodesics with $E=1$ give rise to a new slicing as specified above with flat 3-metric on the slices and a purely radial shift vector
\beq
N^a=-\delta^a_R\epsilon_r Hr\,.
\eeq
This new slicing extends through the above-mentioned coordinate singularity  out to $r\to\infty$.

In the anti de Sitter case we have $\Lambda=-3H^2$ and the metric (\ref{metricgen}) is valid out to infinity in the radial direction.
However, examining Eq.~(\ref{eq:Ur}), one sees that $U_r^2\ge0$ only for $E>1$ and then only within a ball $0<r<\sqrt{E^2-1}/H$ centered at the origin, thus limiting the range of the new coordinates compared to the original ones.
The associated spatial 3-metric is no longer flat as follows from Eq.~(\ref{indmetE}) and the radial shift vector is given by
\beq
N^a=-\delta^a_R\epsilon_r \sqrt{E^2-1-H^2r^2}\,.
\eeq

Painlev\'e-Gullstrand coordinates in the de Sitter spacetime have been introduced by Parikh \cite{parikh} to study tunneling processes across the cosmological horizon.
The same problem has been investigated in the Schwarzschild-anti de Sitter spacetime in Ref. \cite{hemming}, but using a non-geodesic slicing.

\section{The Kerr spacetime}

Now consider the Kerr spacetime with its metric written in Boyer-Lindquist coordinates
\begin{eqnarray}
\label{eq:met}
\rmd s^2 &=& -\left(1-\frac{2Mr}{\Sigma}\right)\rmd t^2 -\frac{4aMr}{\Sigma}\sin^2\theta\rmd t\rmd\phi+ \frac{\Sigma}{\Delta}\rmd r^2 +\Sigma\rmd \theta^2\nonumber\\
&&+\frac{(r^2+a^2)^2-\Delta a^2\sin^2\theta}{\Sigma}\sin^2 \theta \rmd \phi^2\,,
\end{eqnarray}
where $\Delta=r^2-2Mr+a^2$ and $\Sigma=r^2+a^2\cos^2\theta$; here $a$ and $M$ are the specific angular momentum and total mass  characterizing the spacetime. The event horizons are located at $r_\pm=M\pm\sqrt{M^2-a^2}$. 
In this case we have for timelike geodesics \cite{MTW}
\beq
S=\frac12 \mu^2 \lambda -Et +L\phi +S_r(r) +S_\theta(\theta)\,,
\eeq
with
\beq
S_r=\epsilon_r \int \frac{\sqrt{R}}{\Delta} \rmd r\,,\quad 
S_\theta=\epsilon_\theta \int \sqrt{\Theta} \rmd \theta\,,
\eeq
where $\epsilon_r=\pm1$ and $\epsilon_\theta=\pm1$ are sign indicators, and
\begin{eqnarray}
\label{defsvarie}
P&=& E(r^2+a^2)-La\,,\quad 
B= L-aE \sin^2\theta\,, \quad 
R= P^2-\Delta (\mu^2 r^2+K)\,,\nonumber\\
\Theta&=&Q-\cos^2 \theta\left[a^2(\mu^2-E^2)+\frac{L^2}{\sin^2 \theta} \right]\,,\quad 
Q= K-(L-aE)^2\,.
\end{eqnarray}
We set $\mu=1$ so that
\beq
-\rmd T = U^\flat=-E\rmd t+\epsilon_r\frac{\sqrt{R(r)}}{\Delta}\rmd  r+\epsilon_\theta\sqrt{\Theta(\theta)}\rmd  \theta+L\rmd \phi
\eeq
with $R(r)$ and $\Theta(\theta)$ now given by Eq.~(\ref{defsvarie}) with $\mu=1$.
For completeness, we also list below the contravariant components of $U$:
\begin{eqnarray}
U^t&=& \frac{1}{\Sigma}\left[aB+\frac{(r^2+a^2)}{\Delta}P\right]\,,\quad 
U^r=\epsilon_r \frac{1}{\Sigma}\sqrt{R}\,,\nonumber \\
U^\theta&=&\epsilon_\theta \frac{1}{\Sigma}\sqrt{\Theta}\,,\quad 
U^\phi= \frac{1}{\Sigma}\left[\frac{B}{\sin^2\theta}+\frac{a}{\Delta}P\right]\,.
\end{eqnarray}

The induced metric Eq.~(\ref{eq:gamma}) on the hypersurfaces of the foliation $T=\hbox{\it const}$  is obtained by replacing $\rmd t$ by
\beq
\rmd t=\frac{1}{E}\,U_a \rmd x^a
    =\epsilon_r\frac{\sqrt{R(r)}}{E\Delta}\rmd  r+\epsilon_\theta\frac{\sqrt{\Theta(\theta)}}{E}\rmd  \theta+\frac{L}{E}\rmd \phi
\eeq
in Eq.~(\ref{eq:met}), where $x^a=\{r,\theta,\phi\}$. This
depends only on the spatial coordinates $x^a$ and on the three constants of the motion $E,L,K$.

One is free to pick any choice of the three parameters $E,L,K$ to determine a geodesic slicing. If we choose them so that the family includes equatorial geodesics, we must impose first the condition $Q=0$ so that on the equatorial plane $\theta=\pi/2$ one has $U_\theta=0$. However, off the equatorial plane $U_\theta^2$ will be negative if $a\neq0 $ and $E<1$, and will change sign if $E>1$, $L\neq0$, limiting the slicing to exclude a range of $\theta$ values around the polar axis that these geodesics cannot reach. For the slicing to be valid everywhere outside the horizon, one must therefore have $L=0$ and $E=1$, leaving $K=a^2$.
This limits the zero angular momentum geodesics to be at rest at spatial infinity. The infalling geodesics have $\epsilon_r=-1$.
These are the world lines of the Painlev\'e-Gullstrand observers.  

The $T=\hbox{\it const}$ hypersurfaces then correspond to
\beq
\rmd t=U_r \rmd r=\epsilon_r \frac{\sqrt{2Mr(r^2+a^2)}}{\Delta}\rmd r\,,
\eeq
so that the induced metric is given by
\begin{eqnarray}
\label{eq:3met}
{}^{(3)}\rmd s^2
&=& \frac{\Sigma}{\Delta}\rmd r^2+\Sigma\rmd \theta^2+
\frac{\Delta\Sigma\sin^2\theta}{\Sigma -2Mr}\rmd \phi^2  \nonumber\\
&& -\left(1-\frac{2Mr}{\Sigma}\right)\left( U_r\rmd r +\frac{2aMr\sin^2\theta}{\Sigma -2Mr}  \rmd\phi\right)^2\,.
\end{eqnarray}

A direct calculation shows that the metric determinant $\gamma$ and the associated Ricci scalar ${}^{(3)}R$ evaluate to
\beq
\sqrt{\gamma}=\Sigma \sin \theta, \quad
 {}^{(3)}R=\frac{2a^2Mr(3\cos^2\theta -1)}{\Sigma^3} \,,
\eeq
while the trace of the extrinsic curvature is given by
\beq
{\rm Tr}(K)=\frac{\epsilon_r}{2\Sigma} \sqrt{\frac{2M}{r}}\frac{(3r^2+a^2)}{\sqrt{r^2+a^2}}\,.
\eeq
The geometry associated with Painlev\'e-Gullstrand observers in the Kerr spacetime is not intrinsically or  extrinsically flat,  nor is it conformally flat \cite{prigar,kroon1,kroon2}.

Finally a new global coordinate system for Kerr spacetime is given by $(X^\alpha)=(T,R,\Theta,\Phi)$ with $R=r$ and $\Theta=\theta$ unchanged and $T=t+f(r)$ and $\Phi=\phi+h(r)$ such that
\begin{eqnarray}
-\rmd T&=&-\rmd t+U_r\rmd r=U^\flat\,,
  \qquad \rmd R=\rmd r\,, \qquad \rmd\Theta=\rmd\theta\,,
\nonumber\\
\rmd\Phi&=&\rmd \phi +\frac{g^{t\phi}}{g^{rr}U_r}\rmd r=\rmd \phi -\frac{a}{r^2+a^2}U_r\rmd r\,,
\end{eqnarray}
with inverse relations
\begin{eqnarray}
\partial_T&=&\partial_t\,, 
  \qquad \partial_\Theta=\partial_\theta\,, \qquad \partial_\Phi=\partial_\phi\,,
\nonumber\\
\partial_R&=&U_r\left(\partial_t+\frac{a}{r^2+a^2}\partial_\phi\right)+\partial_r\,. 
\end{eqnarray}
The new azimuthal coordinate is of the form $\Phi=\phi+F(r)$, where 
$F'(r)=g^{t\phi}/({g^{rr}U_r})$ is a function only of $r$ since 
$U_r$ is such a function because of the separability condition while
the ratio
\beq
\frac{g^{t\phi}}{g^{rr}}=-\frac{2Mar}{\Delta^2}
\eeq
serendipidously happens to be independent of $\theta$. 
The introduction of $\Phi$ leads to
a zero shift vector component along the azimuthal direction in the new coordinate system
\beq
N^\Phi =
g^{T\Phi}=0\,,
\eeq
thus aligning the new radial coordinate lines with the geodesic observers by incorporating their azimuthal motion into the new azimuthal coordinate.
One finds that
$
g^{\Phi\Phi}=({(r^2+a^2)\sin^2\theta})^{-1}
$,
so the spatial 1-form
\beq
\bar U^\flat=\sqrt{r^2+a^2}\, \sin \theta \rmd \Phi
\eeq
is a unit 1-form  orthogonal to $U^\flat$.
The Kerr metric in this new set of coordinates $X^\alpha$ has been given by Doran \cite{doran} (see also \cite{natario}).
It is again Eq.~(\ref{newmet}), with unit lapse factor and the shift vector aligned with the radial direction, i.e.,
\beq
N^a=\delta^a_R\frac{(1+g_{tt})(r^2+a^2)}{U_r[g_{tt}(r^2+a^2)+ag_{t\phi}]}=-\delta^a_R\epsilon_r\frac{\sqrt{2Mr(r^2+a^2)}}{\Sigma}\,.
\eeq
The nonvanishing components of the spatial metric are instead given by
\begin{eqnarray}
\gamma_{RR}&=&\frac{1+g_{tt}}{(N^R)^2}=\frac{\Sigma}{r^2+a^2}\,, \quad
\gamma_{R\Phi}=\frac{g_{t\phi}}{N^R}=\epsilon_r a\sin^2\theta\sqrt{\frac{2Mr}{r^2+a^2}}\,, \nonumber\\
\gamma_{\theta\theta}&=&g_{\theta\theta}\,, \quad
\gamma_{\Phi\Phi}=g_{\phi\phi}\,.
\end{eqnarray}
The Doran form of the metric (\ref{newmet}) is obtained simply by completing the square on the radial term as follows
\begin{eqnarray}\label{completesquare}
\rmd s^2&=&-\rmd T^2+\gamma_{RR}(\rmd r+N^R \rmd T)^2+2\gamma_{R\Phi}(\rmd r+N^R \rmd T)\rmd \Phi\nonumber\\
&&+\gamma_{\theta\theta}\rmd\theta^2+\gamma_{\Phi\Phi}\rmd\Phi^2\nonumber\\
&=&-\rmd T^2+\gamma_{RR}\left[\rmd r+N^R \rmd T+\frac{\gamma_{R\Phi}}{\gamma_{RR}}\rmd \Phi\right]^2\nonumber\\
&&+\gamma_{\theta\theta}\rmd\theta^2+\left(\gamma_{\Phi\Phi}-\frac{\gamma_{R\Phi}^2}{\gamma_{RR}}\right)\rmd\Phi^2\,,
\end{eqnarray}
with
\beq
\gamma_{\Phi\Phi}-\frac{\gamma_{R\Phi}^2}{\gamma_{RR}}=(r^2+a^2)\sin^2\theta\,, \qquad
\frac{\gamma_{R\Phi}}{\gamma_{RR}}=\epsilon_r a\sin^2\theta\frac{\sqrt{2Mr(r^2+a^2)}}{\Sigma}\,.
\eeq
Recalling that $U^\flat=-\rmd T$ and $\bar U^\flat=\sqrt{r^2+a^2}\, \sin \theta \rmd \Phi$ the final form of the metric is then
\beq
\rmd s^2=-(U^\flat)^2+(\bar U^\flat)^2+\gamma_{RR}\left[\rmd r+N^R \rmd T+\frac{\gamma_{R\Phi}}{\gamma_{RR}}\rmd \Phi\right]^2+\gamma_{\theta\theta}\rmd\theta^2\,,
\eeq
identifying in this way a natural orthonormal frame adapted to $U$ introduced by Doran \cite{doran}.
\begin{eqnarray}
\omega^0&=-U^\flat\,,\qquad&
\omega^1=\sqrt{\gamma_{RR}}\left[\rmd r+N^R \rmd T+\frac{\gamma_{R\Phi}}{\gamma_{RR}}\rmd \Phi\right]\,,\nonumber\\
\omega^2&=\sqrt{\gamma_{\theta\theta}}\rmd\theta\,,\qquad&
\omega^3=\bar U^\flat\,.
\end{eqnarray}

In the special case of the Schwarzschild spacetime, the induced metric reduces to the flat spatial metric as shown above, and the single nonzero shift component in the new coordinate system reduces to $N^R=-\epsilon_r\sqrt{{2M}/{r}}$.
This entire discussion can be repeated for the more general Kerr-Newman family of spacetimes with similar results.

\section{The G\"odel spacetime}

The G\"odel spacetime \cite{godel,Hawell} is a stationary and axisymmetric solution of the Einstein's equations whose metric expressed in cylindrical coordinates $(t,r,\phi,z)$ is
\begin{eqnarray}
\rmd s^2&=&\frac{2}{\omega^2}\left[-\rmd t^2 +\rmd r^2 +\sinh^2 r(1-\sinh^2 r)\rmd \phi^2 \right.
\nonumber\\
&&\qquad \left. +2\sqrt{2}\sinh^2r \rmd t \rmd \phi  +\rmd z^2\right]\,.
\end{eqnarray}
Its matter source is a constant dust energy density $\rho$ (i.e., stress-energy tensor  $T=\rho u\otimes u$), with unit 4-velocity $u=(\omega/\sqrt{2})\partial_t$ aligned with the time coordinate lines and cosmological constant $\Lambda=-\omega^2=-4\pi\rho$, where  $\omega>0$ is the constant rotation parameter chosen to be positive to describe an intrinsic counterclockwise rotation of the spacetime around the $z$-axis.

Define the  radius $r_h$ where $ g_{\phi\phi}=0$ (the $\phi$ coordinate circles are null here, then timelike for larger radii) by
\beq
r_h=\ln (1+\sqrt{2})\approx 0.88137\,,\qquad 
\sinh r_h=1\,,\quad 
\cosh r_h=\sqrt{2}\,.
\eeq 
Unlike the Kerr case, the time lines here are always timelike geodesics, so the coordinates are valid at all radii, but the spacelike time coordinate hypersurfaces used to introduce fiducial observers along their normal direction turn timelike beyond $r_h$, so this represents an observer horizon for this family. The 4-velocity of these fiducial observers is
\beq
    n= \frac{\omega}{\sqrt{2}} \frac{\sqrt{1-\sinh^2r}}{\cosh r} \left(\partial_t -\frac{\sqrt{2}}{1-\sinh^2 r} \partial_\phi\right) \,, \quad r<r_h\,.
\eeq
This observer horizon is similar to the one which occurs for uniformly rotating cylindrical coordinates in Minkowski spacetime where the angular speed of the corotating observer family grows to the speed of light at the light cylinder which terminates their existence.

The covariant representation of the matter 4-velocity $u$ is 
\beq
\label{uflatgodel}
u^\flat=-\frac{\sqrt{2}}{\omega}\left(\rmd t -\sqrt{2}\sinh r \rmd \phi \right)\,,
\eeq 
which is easily seen to be geodesic and shear-free but which has constant nonzero vorticity
\beq
\omega(u)=\frac{\omega}{\sqrt{2}}\partial_z\,,\qquad \omega(u)^\flat= \sqrt{2}\, \rmd z\,.
\eeq
Due to the existence of the three Killing vectors fields $\partial_t$, $\partial_\phi$ and $\partial_z$, the geodesic equations are separable and the covariant 4-velocity of a general timelike geodesic has the following separated form (setting already $\mu=1$ for a proper time parametrization)
\beq
U^\flat=-E\rmd t +L\rmd \phi +b\rmd z +\partial_r S_r(r)\rmd r\,,
\eeq 
where from the normalization condition $U^\alpha U_\alpha=-1$ one then finds
\beq\label{eq:c2}
\left(\frac{\rmd S_r(r)}{\rmd r}\right)^2=U_r^2=\frac{1}{\omega^2 \sinh^2 r}\left[{\mathcal A}\cosh^2 r + {\mathcal B}+\frac{{\mathcal C}}{\cosh^2 r}\right]
\eeq
with the constants ${\mathcal A}$, ${\mathcal B}$ and ${\mathcal C}$ given by
\begin{eqnarray}
{\mathcal A}&=& -(\omega^2b^2+2+\omega^2E^2)\,,\nonumber\\
{\mathcal B}&=& \omega^2b^2+2+3\omega^2E^2+2\sqrt{2}\omega^2LE\,,\nonumber\\
{\mathcal C}&=& -\omega^2(\sqrt{2}E+L)^2\,,
\end{eqnarray}
with ${\mathcal A}+{\mathcal B}+{\mathcal C}=-\omega^2 L^2$.
The 4-velocity vector is
\begin{eqnarray}
   U &=& \frac{\omega^2}{2\cosh^2 r} \left[ \sqrt{2} L + E (2-\cosh^2 r) \right] \,\partial_t 
       +\frac{\omega^2}{2} U_r \partial_r 
\nonumber\\
&& \quad
- \frac{\omega^2}{2 \sinh^2 r \cosh^2 r} \left( E \sqrt{2} \sinh^2 r - L\right) \,\partial_\phi
       + \frac{\omega^2}{2} b \partial_z 
\,.
\end{eqnarray}

By setting $b=0$ we can avoid the unnecessary complication of additional translational motion along the axis of cylindrical symmetry, so that
the above normalization condition simplifies to
\beq\label{eq:U2}
U_r^2 = \left(\frac{2}{\omega^2} \frac{\rmd r}{\rmd \tau}\right)^2
=\frac{\kappa(r) }{\omega^2}(\omega E-V_+)(\omega E-V_-)\,,
\eeq
where the overall coefficient
$\kappa (r)=({2-\cosh^2r})/{\cosh^2 r}$
is positive when $0<r<r_h$ and negative when $r>r_h$
and the effective potentials $V_\pm$ are given by
\beq
V_\pm = \frac{\sqrt{2}\omega L \pm \coth r \sqrt{\omega^2 L^2-2\sinh^2 r (\cosh^2 r-2)}}{\cosh^2 r-2}\,.
\eeq
Noting that only the combinations $L\omega$ and $E\omega$ occur in these formulas for $V_\pm$ and $U_r$, we set
 $\omega=1$ (measuring $L,E$ in units of $1/\omega$).
$V_\pm$ are real only when $0<r<r_*$, where
$\sinh r_*={\sqrt{1+\sqrt{1+2L^2}}}/{\sqrt{2}}$.
At $r=r_*$,  the two potentials meet $V_+=V_-$  and assume the common value
$\lim_{r\to r_*^-}V_\pm ={\sqrt{2}}(1+\sqrt{1+2L^2})/{L} $.
Since $u=\omega/\sqrt{2} \partial_t$ is always a future-pointing timelike 4-vector, and the sign-reversed inner product of any two future-pointing unit vectors must be greater than 1, then
\beq
   -u_\alpha U^\alpha = \frac{\omega}{\sqrt{2}} \partial_t \cdot U = \frac{\omega E}{\sqrt{2}}\ge 1 \rightarrow E\ge \frac{\sqrt{2}} {\omega} \,,
\eeq
so only positive values of the energy $E$ greater than $\sqrt{2}/\omega$ are allowed.

Fig.~1  
shows typical profiles for the effective potentials. Increasing $L>0$ pulls the asymmetric potential well  in (a)  farther and farther to the right past the horizon radius $r_h$, with the maximum value $r_*$ a slowly increasing function of $L$. Thus for geodesics rotating in the same sense as the matter content of the spacetime ($L>0$), the corresponding time slices remain spacelike farther into the region $r>r_h$ containing timelike azimuthal coordinate circles as the angular momentum increases.

\def\SC{0.25}
\begin{figure}
\typeout{*** EPS figure 1}
\begin{center}
$\begin{array}{cc}
\includegraphics[scale=\SC]{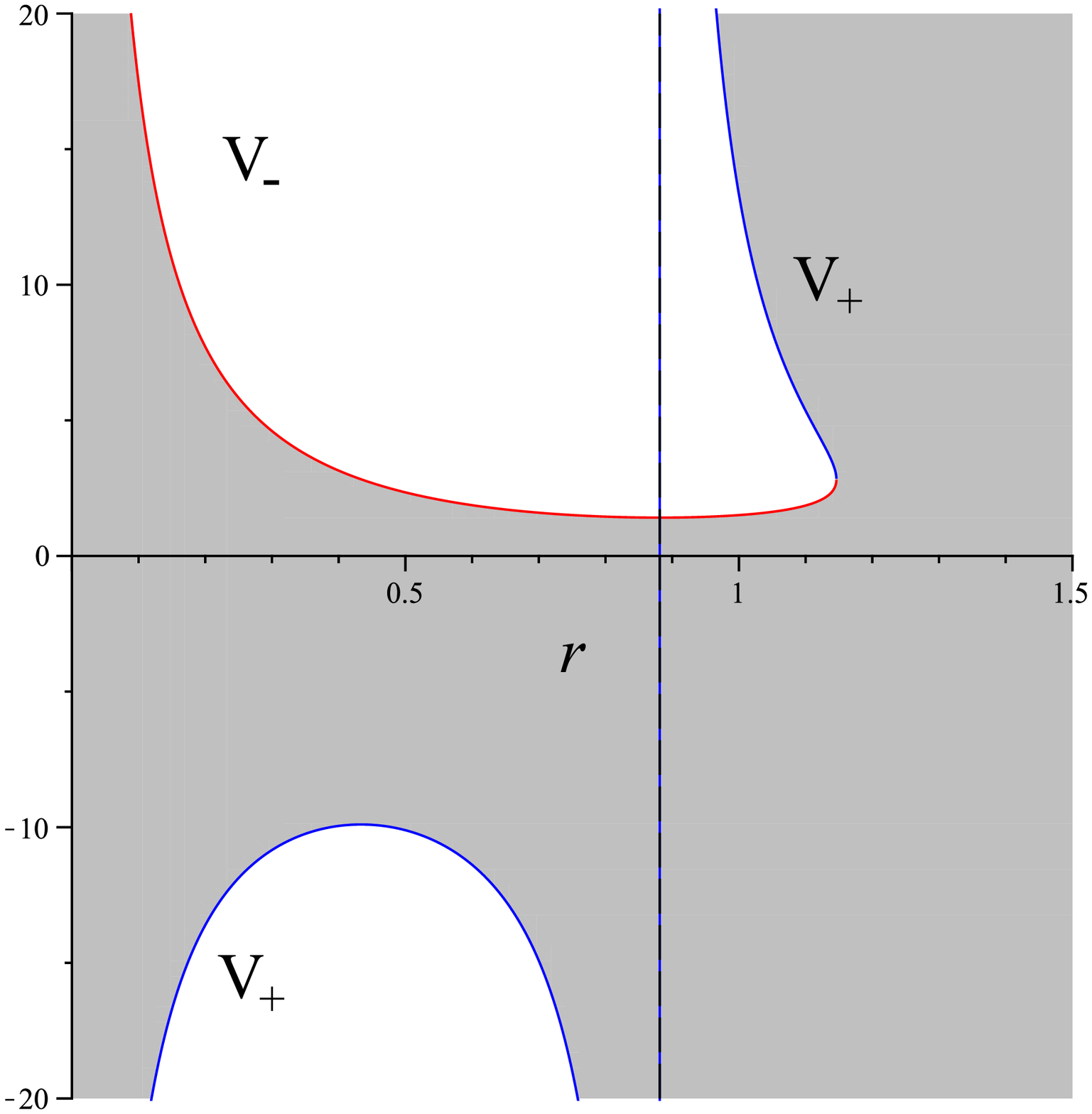}&\quad
\includegraphics[scale=\SC]{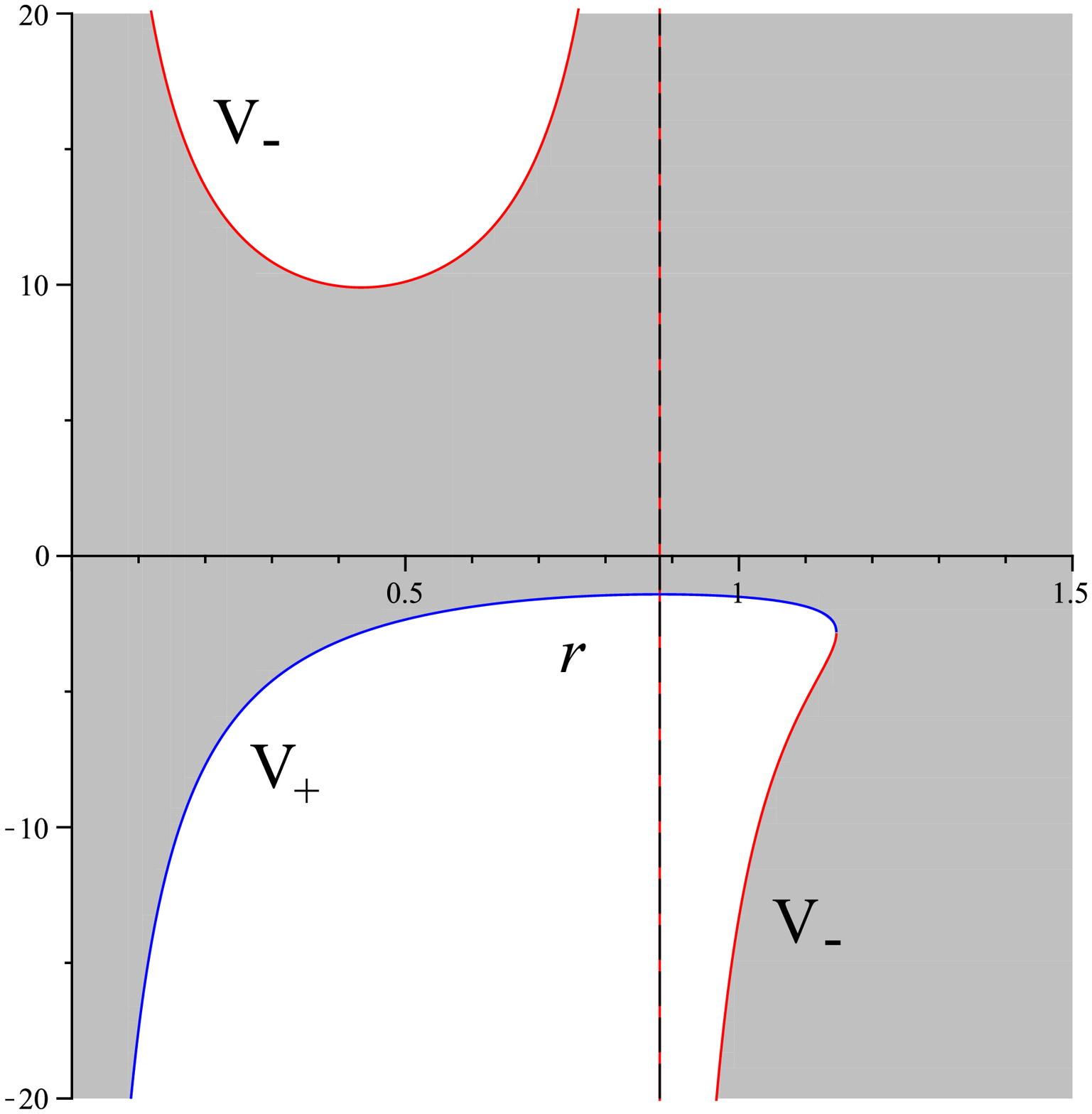}\\[.4cm]
\quad\mbox{(a)}\quad &\quad \mbox{(b)}\\
\end{array}$\\[.6cm]
\includegraphics[scale=\SC]{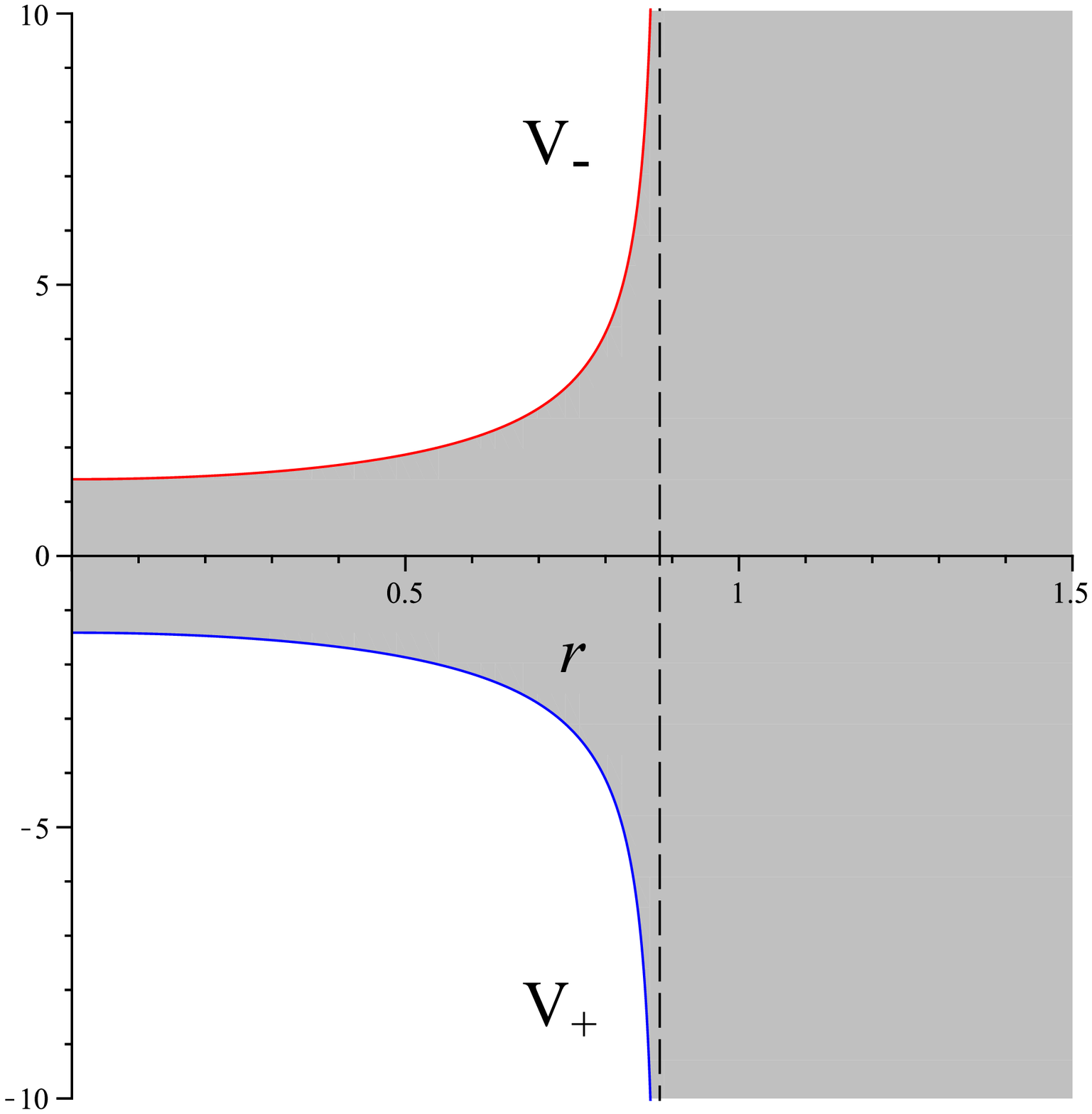}\\[.4cm]
\mbox{(c)}
\end{center}
\label{fig:1}
\caption{The behavior of the effective potentials for radial motion as functions of $r$ is shown for fixed values of $L=[2,-2,0]$ in Figs.~(a) to (c) respectively, having set $\omega=1$. For $U_r^2\ge0$, $E$ must lie above or below both potentials if $0\le r< r_h$ but between them for $ r_h\le r\le r_*$, which are the unshaded regions of the plane.
Changing the sign of $L$ reflects (a) into (b),
while the case $L=0$ in (c) has an obvious reflection symmetry across the horizontal axis, but since $E>\sqrt{2}$, the lower half planes are forbidden regions (corresponding to past-directed 4-velocities). 
Thus horizontal lines representing energy levels in the white region of the upper half plane describe the allowed radial motion.
}
\end{figure}

Radial turning points occur at the zeros $r_\pm$ of (\ref{eq:c2}) or (\ref{eq:U2}) where the energy level intersects the effective potential graph at $E=V_+$ or $E=V_-$, namely 
\beq
\cosh^2 r_\pm=\frac{-{\mathcal B}\pm\sqrt{\Delta}}{2{\mathcal A}}\,,
\eeq
where
\beq
\Delta={\mathcal B}^2-4{\mathcal A}{\mathcal C}=( E^2-2)(  E- E_-)(  E-  E_+)\,,
\eeq
and
\beq
 E_\pm =-2\sqrt{2}  L\pm\sqrt{4 L^2+2}\,.
\eeq
If $L>0$ then $E_-<0$, whereas $E_+>0$ for $ L<1/\sqrt{2}$ and is always negative otherwise.
In contrast, if $L<0$ then $E_+>0$, whereas $E_-<0$ for $ L>-1/\sqrt{2}$ and is always positive otherwise.
Finally, for $L=0$ we have $E_\pm=\pm\sqrt{2}$. 

The radial turning points limit the radial range of this family of geodesics, and hence the range of  the new time coordinate defined by their normal hypersurfaces, as in the case of the bound geodesics in the Kerr spacetime.
For nonzero angular momentum $L>0$ aligned with the angular velocity of the spacetime itself, although one introduces a centrifugal barrier around the axis $r=0$ of cylindrical symmetry, one extends the range of the new time coordinate hypersurfaces orthogonal to this family of geodesics through the original coordinate horizon, thus ``penetrating" this artificial horizon used in defining the fiducial observers associated with the time foliation. By increasing $L>0$, this penetration is increased, at the expense of pushing the centrifugal wall near the origin farther to the right, but as one lowers $E$, eventually the penetration radius $r_+$ is decreased to meet the increasing $r_-$ and the interval over which the family of geodesics is defined shrinks to zero width. This effect does not occur for $L<0$ where the geodesics counterrotate with respect to the angular velocity of the spacetime; the centrifugal potential barriers at $r=0$ and $r=r_h$ only shrink the zone of validity of the new coordinates in this case.

Next we can repeat the same subsequent steps as for the Kerr spacetime,  adapting the slicing to a new temporal coordinate $T$  such that 
\beq
\rmd T=E\rmd t -U_r\, \rmd r - L\rmd\phi\,.
\eeq
The spatial metric induced on the $T=\hbox{\it const} $ hypersurfaces is then 
\begin{eqnarray}
\label{3_metricgodel}
{}^{(3)}\rmd  s^2 &=&  \gamma_{rr}\rmd r^2 +  \gamma_{\phi\phi} \rmd \phi^2+2 \gamma_{r\phi} \rmd r \rmd \phi +\gamma_{zz}\rmd z^2\,,
\end{eqnarray}
with the new coordinate components
\begin{eqnarray}
\gamma_{rr}&=& g_{rr}+\frac{g_{tt}}{E^2}U_r^2\,,\quad 
 \gamma_{\phi\phi}=g_{\phi\phi}+g_{tt}\frac{L^2}{E^2}+2g_{t\phi}\frac{L}{E}\,,\nonumber\\
 \gamma_{r\phi}&=&\frac{U_r}{E}\left(g_{t\phi}+g_{tt}\frac{L}{E}\right)\,, \quad 
\gamma_{zz}=g_{zz}\,,
\end{eqnarray}
depending only on $r$.
The spatial geometry is not intrinsically flat, since
$
{}^{(3)}{}R^{r\phi}{}_{ r\phi}=\omega^4 E^2
$
is the only nonvanishing coordinate component of the spatial Riemann tensor,
corresponding to a constant spatial Ricci scalar
${}^{(3)}{}R=2\omega^4E^2$.
However, the Cotton-York tensor \cite{cotton_bob,bindef} is identically zero implying that the spatial metric is conformally flat. The extrinsic curvature is nonzero.

Finally a new coordinate system for the G\"odel spacetime within the horizon radius of this family of geodesics
is given by $(T,R,\Phi,Z)$ with $R=r$ and $Z=z$ unchanged and $T=T(t,r,\phi)$ and $\Phi=\Phi(\phi,r)$ such that
\beq
\label{trasfgod}
\rmd T =E\rmd t-U_r\rmd r-L\rmd\phi\,,\quad
\rmd R=\rmd r\,, \quad
\rmd\Phi =\rmd \phi +{\mathcal F}\rmd r\,, \quad
\rmd Z=\rmd z\,,
\eeq
with inverse relations
\begin{eqnarray}
\partial_T&=&\frac{1}{E}\partial_t\,,\
\partial_\Phi=\frac{L}{E}\partial_t+\partial_\phi\,,\nonumber\\
\partial_Z&=&\partial_z\,,\  
\partial_R=\frac{U_r-L{\mathcal F}}{E}\partial_t-{\mathcal F}\partial_\phi+\partial_r\,,  
\end{eqnarray}
where
\beq
{\mathcal F}=\frac{Eg^{t\phi}-Lg^{\phi\phi}}{g^{rr}U_r}\,.
\eeq
This choice corresponds to aligning the shift vector field with the new radial direction so that
$
N^\Phi =g^{T\Phi}=0
$,
exactly as in the Kerr case. 
One then finds
\beq
g^{\Phi\Phi}
=\frac{\omega^2E^2-2}{2\sinh^2r\cosh^2r \, U_r^2}\,,
\eeq
reintroducing the general value of $\omega$ into the discussion.
Therefore the spatial 1-form $\bar U^\flat=(g^{\Phi\Phi})^{-1/2} \rmd \Phi$ has unit length and is orthogonal to $U$.
The G\"odel metric in this new set of coordinates is then given by Eq.~(\ref{newmet}), with unit lapse factor and the shift vector aligned with the radial direction, i.e.,
\beq
N^a=-\delta^a_R\frac12\omega^2U_r
\,, 
\eeq
and nonvanishing components of the spatial metric given by
\begin{eqnarray}
\gamma_{RR}&=&\frac{4(E^2\omega^2-2)}{E^2U_r^2 \omega^6}\,, \quad
\gamma_{R\Phi}=-\frac{4(E\sqrt{2}\sinh^2 r-L)}{E^2U_r\omega^4}\,, \quad
\gamma_{zz}=g_{zz}=\frac2{\omega^2}\,, \nonumber\\
\gamma_{\Phi\Phi}&=&\frac{2}{\omega^2 E^2} \sinh^2 r (2- \cosh^2 r) (E-W_+)(E-W_-)\,,
\end{eqnarray}
where
\beq
W_\pm=\frac{-L\sqrt{2}\pm |L| \coth r}{(2-\cosh^2 r)}\,.
\eeq

One can also easily diagonalize the new form of the spacetime metric exactly as done in Eq.~(\ref{completesquare}) for the Kerr case leading  to a Doran-like sum of squares representation of the new form of the metric,
namely
\begin{eqnarray}
\rmd s^2&=&-\rmd T^2+\gamma_{RR}(\rmd r+N^R \rmd T)^2+2\gamma_{R\Phi}(\rmd r+N^R \rmd T)\rmd \Phi\nonumber\\
&&+\gamma_{\Phi\Phi}\rmd\Phi^2+\gamma_{zz}\rmd z^2\nonumber\\
&=&-\rmd T^2+\gamma_{RR}\left[\rmd r+N^R \rmd T+\frac{\gamma_{R\Phi}}{\gamma_{RR}}\rmd \Phi\right]^2\nonumber\\
&&+\left(\gamma_{\Phi\Phi}-\frac{\gamma_{R\Phi}^2}{\gamma_{RR}}\right)\rmd\Phi^2+\gamma_{zz}\rmd z^2\,,
\end{eqnarray}
with coefficients
\begin{eqnarray}
\gamma_{\Phi\Phi}-\frac{\gamma_{R\Phi}^2}{\gamma_{RR}}&=&\frac{2\sinh^2r\cosh^2r U_r^2}{(E^2\omega^2-2)}
  = (g^{\Phi\Phi})^{-1}
\,, \nonumber\\
\frac{\gamma_{R\Phi}}{\gamma_{RR}}&=&\frac{\omega^2U_r}{E^2\omega^2-2}(L-\sqrt{2}E\sinh^2r)\,,
\end{eqnarray}
while the new spatial metric determinant is $4\sinh^2 2r/(E^2 \omega^8)>0$.
Recalling that $U^\flat=-\rmd T$ and $\bar U^\flat=(g^{\Phi\Phi})^{-1/2}\rmd \Phi$ we get
\beq
\rmd s^2=-(U^\flat)^2+(\bar U^\flat)^2+\gamma_{RR}\left[\rmd r+N^R \rmd T+\frac{\gamma_{R\Phi}}{\gamma_{RR}}\rmd \Phi\right]^2+\gamma_{zz}\rmd z^2\,,
\eeq
identifying in this way a natural orthonormal frame adapted to $U$
\begin{eqnarray}
\omega^0&=-U^\flat\,,\qquad&
\omega^1=\sqrt{\gamma_{RR}}\left[\rmd r+N^R \rmd T+\frac{\gamma_{R\Phi}}{\gamma_{RR}}\rmd \Phi\right]\,,\nonumber\\
\omega^2&=\bar U^\flat\,,\qquad&
\omega^3=\sqrt{\gamma_{zz}}\rmd z\,.
\end{eqnarray}

\section{Concluding remarks}

We have shown how stationary spacetimes admitting separable geodesic equations admit a new spacetime slicing orthogonal to a particular family of timelike geodesics, corresponding to a unit lapse gauge. For stationary spherically symmetric vacuum spacetimes, an intrinsically flat slicing is possible, as for the Schwarzschild spacetime, reproducible also for the de Sitter spacetime.
In the stationary axisymmetric case of the Kerr and G\"odel spacetimes, the separability also explains the existence of a new  azimuthal coordinate which allows the alignment of the shift vector field with the spherical radial or cylindrical radial direction respectively. In the Kerr spacetime this leads to the usual Painlev\'e-Gullstrand slicing and adapted coordinate system, while in the G\"odel case it leads to a new analogous form of the metric.

\section*{Acknowledgments}

All authors thank ICRANet for support. 
DB acknowledges O. Semer\'ak for useful discussion.

\end{document}